\documentclass[a4paper]{article}

\usepackage[margin=1.18in]{geometry} %

\usepackage{booktabs}
\usepackage[T1]{fontenc}
\usepackage{hyperref}
\hypersetup{
    unicode,
    hidelinks,
    colorlinks,
    urlcolor=blue, 
    linkcolor=blue,
    citecolor=blue,
    pdfauthor={Erica Forzinetti, Marco L. Della Vedova, Stefano Pasta, Milena Santerini},
    pdftitle={Indicators for characterising online hate speech and its automatic detection},
    pdfkeywords={hate speech, machine learning, social networks, detection},
}

\usepackage[numbers,square]{natbib}
\usepackage{graphicx, color}
\usepackage{url}
\usepackage{tabularx}

\title{\textbf{Indicators for characterising online hate speech and its automatic detection}}

\author{
    Erica~Forzinetti$^1$ 
    \and Marco~L.~Della~Vedova$^{1,2,}$\thanks{Corresponding author: marco.dellavedova@chalmers.se} 
    \and Stefano~Pasta$^1$
    \and Milena~Santerini$^1$
}

\date{
    $^1$Osservatorio sull'odio on line - Mediavox\\
    Università Cattolica del Sacro Cuore \\ 
    Milano, Italy \\[2ex]
    $^2$Chalmers University of Technology \\ 
    Gothenburg, Sweden\\[2ex]%
    \today
}

\begin{document}
\maketitle

\begin{abstract}
We examined four case studies in the context of hate speech on Twitter in Italian from 2019 to 2020, aiming at comparing the classification of the 3,600 tweets made by expert pedagogists with the automatic classification made by machine learning algorithms. 
Pedagogists used a novel classification scheme based on seven indicators that characterize hate. 
These indicators are:
the content is public, it affects a target group, it contains hate speech in explicit verbal form, it will not redeem, it has intention to harm, it can have a possible violent response, it incites hatred and violence. 
The case studies refer to Jews, Muslims, Roma, and immigrants target groups. 
We find that not all the types of hateful content are equally detectable by the machine learning algorithms that we considered.
In particular, algorithms perform better in identifying tweets that incite hatred and violence, and those that can have possible violent response.
\\[2ex]%
\end{abstract}

\section{Introduction}
In the public language, speeches and narratives of the infosphere and \textit{onlife} society~\cite{floridi2014}, i.e., marked by the continuity of cross-references between online and offline, discursive networks have been set up and they are able to build new paradigms of hatred~\cite{pasta2018,santerini2021}. 
Invectives, insults, offences, and rhetorics of aversion follow human relations and conflicts throughout history, but a singular reflection on this phenomenon is underway today, increasingly perceived as a vital issue for the stability of democracy and social life. 
The spread of Web 2.0 has opened up new dimensions to hate speech, transforming its structure and syntax, but first of all its meaning and motivations. Most of this communication, which is spreading in liquid, unstructured and trivialised form, takes place under the banner of emotions, which orient and direct our minds intelligently but also quickly and spontaneously~\cite{damasio1994,nussbaum2008,wallace2015}.
From a legal perspective, \textit{hate speech} has no clear definition~\cite{ziccardi2016,ziccardi2020,faloppa2020}. 
For the Council of Europe, it means ``all forms of expression that spread, incite, promote or justify racial hatred, xenophobia, anti-Semitism or other forms of hatred based on intolerance, including: intolerance expressed by aggressive nationalism and ethnocentrism, discrimination and hostility against minorities, migrants and people of immigrant origin''~\cite{eurec1997}.
On the other hand, one could speak of hatred as a feeling, a persistent and systematic attitude of aversion towards another person or group, or as an intrinsic psychological dynamic that may or may not induce violent behavior~\cite{santerini2019}.
Because of its multidimensionality, the phenomenon requires a multidisciplinary approach. From the legal field, the discussion has reached the humanities (sociological, pedagogical, anthropological, philosophical, linguistic, semiotic), converging in the interdisciplinary field of Hate Studies, which brings together scholars, researchers, politicians, communication experts, human rights activists, NGO leaders.
From a pedagogical point of view, education for citizenship is affected by hostile narratives, by the binary vision of the world, divided into us/them, friend/enemy, inside/outside, by the affirmation of processes of choosing a widespread and banal target group, not only in words but also in images. Hate speech is considered a serious threat to social cohesion, considering that the narrow boundary with freedom of expression and the need to preserve this fundamental right sometimes provide a pretext for allowing it to spread without limits. As can be seen from the qualitative-textual and qualitative-motivational analysis and the following conversations with young authors of hate performances, educational reflection has the need to study the phenomenon in greater depth in order to prevent and combat the various forms expressed in the form of online language and images~\cite{pasta2018}.
The research carried out so far in Italy has made it possible to analyze online hate speech in different ways: in some cases, it has been studied from geographical mapping, in others a general quantification has been attempted using only offensive terms to identify online hate and the results indicate that the phenomenon is more circumscribed and limited~\cite{damico2021}.
The linguistic analyses that have developed on the basis of this assumption have made several steps forward in identifying the lexical components of hate speech~\cite{femia2020,ferrini2019}. 
From a methodological point of view, several analyses have also been made, experimenting with different approaches for automatic speech processing (sentiment analysis, text mining...) and on data collection and annotation to capture the different facets of hate speech~\cite{sanguinetti2018}.
Because of the educational implications of the phenomenon, these initial analyses suggest that the detection of online hatred should be combined with a more in-depth study of its characteristics, creating a synergy between computer science and the humanities. 
In fact, it is not only a question of deciding `what is hate', but also of analyzing the many forms in which it is expressed, in order to better identify subjects, targets and ways of expressing it, with a view to developing more effective prevention and counteraction strategies. 
Previous studies have examined and compared the websites of ideological groups from a communications and media use perspective. 
The results indicated that group type was predicted by the type of information presented, the difficulty of becoming a member, and the amount of freedom members had on discussion boards.
These findings suggest that characteristics of violent ideological group websites can be used to distinguish them from websites of both nonviolent ideological and non-ideological groups~\cite{byrne2013}.

\section{Methodology}
In this paper, we face the problem of classifying hate speech on Twitter\footnote{Data has been collected in 2021, before the rebranding of Twitter to X.} by combining the socio-educational approach and automatic computer processing. 
Thanks to the support of experts pedagogists we try to identify specific features and indicators that characterize hate contents. 
The data has been collected using the TwitterAPI with specific search queries detailed in the next section. 
The expert annotators, given the text of the tweet, the author, the date of publication, the number of likes and retweets, determined whether the tweet contains hate or not. 
If it was hate content, they assigned the corresponding indicators.
The indicators that determine whether a text contains hate are seven:
\begin{enumerate}
    \item \textbf{Public}: the content can be seen without limitation by users, so it is not a private message in a restricted circle; note that in the specific case under consideration, all tweets are public.
    \item \textbf{Target group - other target}: the hate content affects a target group, or an individual related to that group, or an individual for what he or she represents (e.g., the first nurse vaccinated to hit vaccines); often the targets are minorities or their members, as indicated for example by the definition of hate speech Council of Europe Recommendation (97)20, adopted on October 30, 1997, and the texts of the same institution. This process of election differentiates this content from the behaviors of the cyberbullying spectrum.
    \item \textbf{Hate expression}: hate content contains hate speech in explicit verbal form, i.e., hate words, verbally explicit insults, or speech in which one denies the other person as a person, considering them inferior or attributing negative qualities to them, insulting them, or humiliating them.
    \item \textbf{Will not redeem}: the speech is not interested in changing the mind of the hate victim, but only in insulting and hurting him. It should also be considered that very often the target of hate is not the interlocutor, but it is the object of the speech, so rarely the purpose of the author of hate speech is to "redeem" the target, also because the most structured forms of hate do not allow, precisely, the redemption because the target is hated "as such" (Jew, foreigner, black, Roma ...) regardless of his individual behavior.
    \item \textbf{Intention to harm}: the content is marked by an intentionality of the author to hit the victim (individual or target). Even in this case it should be noted that often the target does not coincide with the interlocutor but rather with the object of the speech, then often takes the form of speeches with the intentionality to produce hostility towards the target.
    \item \textbf{Possible violent response}: the tones of speech are marked by violence or intensity such that, following the typical mechanisms of toxic disinhibition on the Web, they easily and quickly allow for a response of incitement to hatred and violence.
    \item \textbf{Inciting hatred and violence}: hate contents that explicitly and directly incite hatred and violence, in which the author aims to enlarge the co-producers of hate speech. In these cases, online hate is a particularly fertile background for offline hate actions and hate crimes.
\end{enumerate}
These seven characteristics has been identified as the ``Spectrum of Online Hate Indicators'', in part to address a definitional ambiguity of the concept of hate, and are presented in increasing intensity. 
The proposal draws from the didactics of the Shoah the interpretive tool of the Anti-Defamation League's ``Pyramid of Hate'', which interprets the extreme outcome of hate as thresholds that are crossed and taboos that are broken down.
However, although the spectrum of indicators of online hate indicates a progression of intensity, the sequence of indicators should not be understood rigidly, and therefore one may find hate content that has some indicators but not others.
In any case, the presence of multiple indicators on the spectrum, and particularly those of higher intensity, detects content with potentially very dangerous effects in the communication ecosystem.

\section{Case studies and data collection}
This analysis has been tested on four datasets referring to four different target groups: Jews, Muslims, Roma and immigrants. 
The first dataset, referred to Jews, is composed by 160,646 Italian tweets published from 2019, September 1st to 2020, May 31st. 
The search query used to download the tweets is 
\begin{center} \it
ebrei OR Soros OR Israele OR sionismo OR sionista
\end{center}
(transl. Jews OR Soros OR Israel OR Zionist OR Zionism).
Those terms were identified from the working definition of Antisemitism by the International Holocaust Remembrance Alliance, from the international reports of the Observatory on Anti-Semitism of the Centre for Contemporary Jewish Documentation, the Kantor Center for the Study of Contemporary European Jewry, as well as on the basis of the literature on anti-Semitism~\cite{taguieff2015,santerini2019,pasta2022};
in this way, all tweets containing at least one of the words in the query were downloaded. 

The second dataset, referred to Muslims, is composed by 8,299 Italian tweets published from 2019, September 1st to 2020, May 31st. The search query used to download the tweets is 
\begin{center} \it
(islam OR musulman\_ OR mussulman\_) AND 
(radical\_ OR integral\_ OR terroris\_ OR valor\_ OR immigra\_ OR sharia OR velo)
\end{center}
(transl. islam OR musulman\_ OR mussulman\_ AND radical\_ OR integral\_ OR terroris\_ OR valu\_ OR immigra\_ OR sharia OR veil), in order to obtain all the tweets containing at least one lemma identifying the target group (words in the first bracket) and simultaneously with a reference to typical elements of Islamophobia (words in the second bracket). These are, in particular, issues at the basis of the most widespread forms of Islamophobia in Italy, associated with the following ``discourses'': the identification of all Muslims to potential terrorists; anti-Muslim hostility that is combined with xenophobic sentiment against immigrants; a clash of values as a form of reaction to the supposed ``Islamization'' of Italy that confessionalizes the theory of ethnic substitution closely linked to culturalist and differentialist racism, based on the impossibility of coexistence due to an irreconcilable cultural difference (``Muslims are too different, it is impossible to live together''); Islam for essence anti-West, that is, the idea based on the accusation of anti-democracy declined towards a secular defense of subjective rights protected by the Constitution, to which all Muslims in Europe - according to this ``vulgate'' - would not want to adapt, in the name of the alleged superiority of religious laws. 

The third dataset, referred to Roma, is composed by 1,191 Italian tweets published from 2020, June 1st to 2020, December 31st. The search query used to download the tweets is 
\begin{center} \it
(zingar\_ OR rom OR nomad\_) AND (feccia OR ladr\_ OR rapi\_ OR rolex OR bulldozer OR strupr\_ OR violent\_ OR mendica\_ OR elemosin\_)
\end{center}
(transl. gypsy OR roma OR nomad\_ AND scum OR thief OR kidnap\_ OR rolex OR bulldozer OR rap\_ OR violent\_ OR beg\_ OR charity), in order to obtain all the tweets containing at least one lemma identifying the target group (words in the first bracket) and simultaneously "hostile words" or words evocative of specific stereotypes and prejudices towards Roma and Sinti (words in the second bracket)~\cite{pasta2019}. Such lemmas, rooted in the Italian “common sense”, are also confirmed by the literature~\cite{claps2011} and by European reports, such as the Civil society monitoring report on implementation of the national Roma integration strategy in Italy. 

The fourth and last dataset, referred to people with a migration background, is composed by 37,817 Italian tweets published from 2020, June 1st to 2020, December 31st. The search query used to download the tweets is
\begin{center} \it
(migrant\_ OR profug\_ OR stranier\_ OR immigrat\_ OR pover\_ OR negr\_) AND 
(scimmia OR banan\_ OR sbarc\_ OR invasion\_ OR razz\_ OR invas\_ OR stupr\_ OR violent\_)
\end{center}
(transl. migrant\_ OR refuge\_ OR foreign\_ OR immigra\_ OR poor\_ OR nigger\_) AND (monkey OR banan\_ OR land\_ OR invasion\_ OR race\_ OR invad\_ OR rap\_ OR violen\_), the first bracket refers to words connected to the migratory flows towards Italy, while two other words have been inserted to evoke socio-economic relations (``poor'') and the classic image of racism towards blacks (``nigger'')~\cite{santerini2021}; the second bracket refers instead to racist images and discourses, with particular reference to the landings that took place towards the Italian coasts and to what the Council of Europe defined in 2021 as ``Afrophobia'' or ``anti-Black racism''~\cite{councilofeu2021}.

For each of the four datasets, a random sample of 900 tweets has been selected to be manually annotated by experts in each field. The annotators assigned to each tweet a binary label (yes/no) for each of the seven indicators. The resulting dataset of 3,600 annotated tweets is analyzed in what follows.

\section{Results}
\begin{figure}[htb]
    \centering
    \includegraphics[width=.75\linewidth]{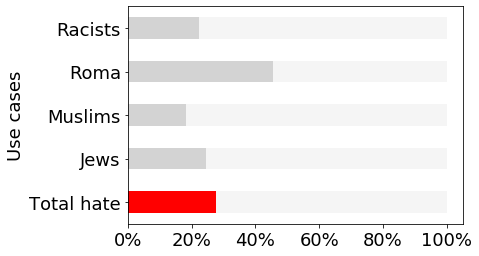}
    \caption{Percentage of hate tweets in the sample dataset}
    \label{fig:totalhate}
\end{figure}

The first result is related to the percentage of hate in the dataset. 
The annotators classified 996 out of 3,600 tweets as containing hate, which corresponds to 27.7\% of the total. 
Fig.~\ref{fig:totalhate} shows the percentage of tweets containing hate divided into the four case studies (grey bars). 
As can be seen, in the Roma dataset more than 40\% of tweets are considered as hateful content, which is a much larger percentage with respect to the other test cases. 
This is probably due to the condition used in the query for downloading the tweets.

Some examples of tweets, which are labelled has hate content, are shown in Table~\ref{tab:example-hate} with the corresponding English translation.

\begin{table}[htb]
    \centering
    \begin{tabularx}{\textwidth}{XX}
    \toprule
    Original text of the tweet & English translation\\
    \midrule
    QUELLA ENORME METASTASI INFILTRATA DI GEORGE SOROS & 
    THAT HUGE INFILTRATED METASTASIS OF GEORGE SOROS\\ \midrule
    Sto dicendo che praticamente tutti i terroristi sono musulmani. Non ti risulta? &
    I'm saying that basically all terrorists are Muslims. Don't you know that? \\ \midrule
    ex zingari...ora sinti..domani rom...ma sempre ladri sono.. 			irrecuperabili!??? & 
    ex Gypsies ... now Sinti ... tomorrow Roma ... but always thieves are ... irretrievable!??? \\ \midrule
    E questi sono i poveretti che salviamo perché hanno bisogno di aiuto?questi sono DELINQUENTI DELLA STESSA RAZZA PDche vengono per rubare spacciare e fottere & 
    And these are the poor people we save because they need help? These are CRIMINALS OF THE SAME KIND who come to steal, deal and fuck\\
    \bottomrule
    \end{tabularx}
    \caption{Examples of tweets containing hate, as classified by annotators\label{tab:example-hate}}
\end{table}

\begin{figure}[htb]
    \centering
    \includegraphics[width=.75\linewidth]{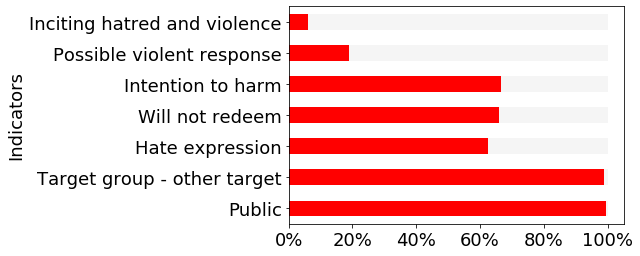}
    \caption{Percentage of tweets characterized by indicators within hate tweets}
    \label{fig:indicatorsperc}
\end{figure}

Once it was established that a tweet contained hate, the second step of the analysis was to check whether the characteristics of the seven indicators were present in the text. The annotators, for each of the 996 hate tweets, labelled which indicator was present or not. The results are shown in Fig.~\ref{fig:indicatorsperc}. All tweets are marked as public and almost all refer to a target group, and this is obviously due to the way in which the dataset was made up. Moreover, only a small percentage of hate tweets (<~20\%) can have a possible violent response and incite hatred and violence.

In order to make a more detailed comparison, the results obtained for each target group (Jews, Muslims, Roma and people with a migration background) were compared using the radarchart technique, a graphical method of visualizing multivariate data. The results are shown in Figure 3. As can be seen, the indicators homogeneously occurred in all four target groups. The most substantial differences can be seen in the dataset referring to Jews, where, unlike the others, a high percentage (more than 90\%) of tweets has intention to harm and will not redeem while only a little percentage can have possible violent response. On the other hand, tweets referred to Muslims have a small percentage (20\%) of hate expression.

\begin{figure}[htb]
    \centering
    \includegraphics[width=.75\linewidth]{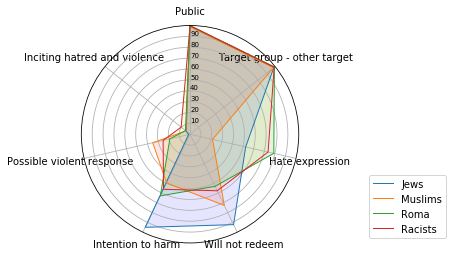}
    \caption{Radarchart of indicators for each target group}
    \label{fig:radarchart}
\end{figure}

The last step of this analysis aims at evaluating automatic hate speech classifiers based on machine learning.
We chose four ML classifiers among the few available open source, pre-trained models, specific for the Italian language.
The test consists of the comparison between the output of the automatic classifiers and the evaluation of the expert annotators, which is assumed to be ground-truth.
The first model, Montanti~\cite{bisconti2020}, was developed for EVALITA-2020~\cite{basile2020,sanguinetti2020}, a periodic evaluation campaign of Natural Language Processing (NLP) and speech tools for the Italian language\footnote{See \url{http://www.evalita.it}}.
The main task of the competition was about the hate speech detection, a binary classification task aimed at determining the presence or the absence of hateful content in the text towards a given target (among immigrants, Muslims and Roma). 
The other three algorithms are available on the website HuggingFace. Dehatebert-mono-italian~\cite{aluru2021} algorithm is used for detecting hate speech in Italian language, while 
POLIticBERT\footnote{Available at \url{https://huggingface.co/unideeplearning/polibert_sa} \label{POLIticBERT}} 
and Neuraly\footnote{Available at \url{https://huggingface.co/neuraly/bert-base-italian-cased-sentiment} \label{Neuraly}} 
performs sentiment analysis on Italian sentences. 
With regard to the latter two algorithms, in order to determine which tweets were to be considered as hate content, an optimal threshold was chosen considering the highest balanced accuracy score. 
Table~\ref{tab:example-algo} shows some of the results obtained from the four algorithms and the label that was assigned by the annotators.

As can be seen in Table~\ref{tab:example-algo}, the first tweet is classified correctly by all four algorithms, probably because there are words like 'dickheads’, ‘terrorists' and ‘Muslims’.
The second tweet still contains negative words as ‘terrorists’, ‘enemy’, and ‘dead’ in fact all the four algorithms classify it as hate content, however, according to the annotators, it is not a question of hatred because the author is criticizing those who hate. 
Conversely, the third tweet contains words as ‘children’ and ‘joking’, in fact all four algorithms classify it as hate-free, but, according to the annotators, it is hate content because it states a prejudice against Roma. 
These last two tweets are clear examples that show how the identification of the context is one of the principal problems for the machine learning algorithms. 
The fourth and fifth tweets can be considered doubtful cases, in fact the fourth is a text that contains real hate against Jews and the fifth, while having negative words, describes a fact that really happened, so it cannot be considered a hateful content. 
Note that in this example POLItic BERT algorithm correctly identified all five tweets, but obviously there are other tweets not classified correctly by this algorithm as well.

\begin{table}[htb]
\centering
\begin{tabular}{@{}lllll@{}}
\toprule
Indicator & \cite{bisconti2020} & \cite{aluru2021} & -\footref{POLIticBERT} & -\footref{Neuraly} \\
\midrule
(i1) Public & 0.62 & 0.65 & 0.65 & 0.62 \\
(i2) Target group - other target & 0.62 & 0.65 & 0.64 & 0.62 \\
(i3) Hate expression & 0.62 & 0.66 & 0.66 & 0.66 \\
(i4) Will not redeem & 0.61 & 0.64 & 0.65 & 0.61 \\
(i5) Intention to harm & 0.60 & 0.63 & 0.65 & 0.61 \\
(i6) Possible violent response & 0.70 & 0.68 & 0.68 & 0.65 \\
(i7) Inciting hatred and violence & 0.63 & 0.70 & 0.69 & 0.69 \\ \midrule
\textbf{Overall} & 0.62 & 0.65 & 0.69 & 0.69\\ 
\bottomrule
\end{tabular}
\caption{Balanced accuracy score of the four ML algorithms\label{tab:accuracy}}
\end{table}

The last analysis is about the performance of automatic classifiers on the seven hate indicators.
Table~\ref{tab:accuracy} shows the balanced accuracy score for each algorithm for each indicator. 
The balanced accuracy score is a standard performance metric used when the dataset is not balanced, i.e., when the number of observations with a label is very different to the number of observations with another label, as it happens in our case.
As can be seen, all the algorithms have the highest balanced accuracy score for ‘Possible violent response’ and ‘Inciting hatred and violence’ indicators. 
In this regard, Table~\ref{tab:possible_violent} contains anonymized examples of tweets which are labelled as ‘Possible violent response’ and/or ‘Inciting hatred and violence’.

As shown in Table~\ref{tab:possible_violent}, the text of these tweets contains many slang words (i.e., ‘assholes’, ‘filthy scum’, ‘jail’, ‘disease’), which can be easily detected by the algorithm.
From the first two examples, it can be seen how these are real hate contents.
Both were classified by the annotators as ‘Inciting hate and violence’ and ‘Possible violent response’, in fact they contain strong and violent hatred and there is a risk that these will result in offline hate. 
Moreover, in the third tweet there is not a violent tone (in fact the label for ‘Possible violent response’ is~0) but it could be a background for offline hate actions and hate crimes (‘Inciting hate and violence’ has label~1). 
On the other hand, the fourth tweet has a rough tone (‘Possible violent response’ is~1) but, according to the annotators, the intent to spread hate is not present (‘Inciting hate and violence’ has label~0).

\section{Discussion and conclusion}
One of the most important factors to emphasize before drawing conclusions from these analyses is that the high percentage of hatred present in the dataset (27,7\%) is due to the way in which the queries to download the tweets were created, in particular for the datasets referred to Muslims, Roma and people with a migration background. In this study, the datasets were made up with all those tweets that contained simultaneously "hostile words" or words evocative of specific stereotypes and prejudices towards.
At the end of this study, one of the positive aspects that could be found is a low percentage of tweets classified as ‘Inciting hate and violence’ and ‘Possible violent response’ by the annotators (Figure~\ref{fig:indicatorsperc}).  This clearly means that the tweets downloaded from the web don't allow to have a hate response easily and quickly (possible violent response) and are not a fertile background for offline hate actions or hate crimes (inciting hatred and violence). Most of the hate found online is not to be considered dangerous, it is often an end in itself. This result is in line with studies already done, since it has been proved that the websites of violent ideological groups that use more fear appeals were less interactive and less credible than non-ideological groups and nonviolent ideological groups~\cite{dunbar2014}. 
However, not having downloaded real-time data, it may be that Twitter's algorithms have promptly eliminated more dangerous hate contents. 
This can also be noticed from the results shown in Table~\ref{tab:accuracy} about the balanced accuracy scores. 
All four algorithms used in this study seem to perform better in identifying tweets classified as ‘Inciting hate and violence’ and ‘Possible violent response’. 
This means that machine learning algorithms are able to better identify the tweets considered most dangerous by the annotators. 
On the other hand, one of the biggest weaknesses of machine learning algorithms is about the context detection. 
Too often it happens that the same words used in different contexts have different meanings. 
This is the main reason why automatic online hate speech detection still has many drawbacks. 
The main problem causing the low reliability of these methods is due to the extremely unbalanced presence of real hateful content in the datasets which are analyzed and the lack of distinctive and discriminatory features.

\begin{table}[htbp]
\centering
\begin{tabularx}{\linewidth}{XXc|cccc}
\toprule
Original text of the tweet & English translation & Hate annot. & \cite{bisconti2020} & \cite{aluru2021} & -\footref{POLIticBERT} & -\footref{Neuraly} \\ 
\midrule
Minoranza forse, però grandissime teste di cazzo che hanno e stanno facendo guerre e attentati in nome di Allah! E guarda caso tutti i terroristi sono musulmani. & A minority perhaps, but huge dickheads who have and are making wars and attacks in the name of Allah! And it just so happens that all terrorists are Muslims. & 1 & 1 & 1 & 1 & 1 \\ \midrule
Mah, passata la moda dei musulmani tutti terroristi, ora tocca agli ebrei, devono trovarsi un nemico, altrimenti non sanno come giustificare la propria esistenza, sono morti & Well, after the fashion for Muslims to be all terrorists, now it's the turn of the Jews, they have to find an enemy, otherwise they don't know how to justify their existence, they're dead. & 0 & 1 & 1 & 1 & 1 \\ \midrule
@user Da piccolo mia mamma diceva" stai lontano dagli zingari perché rapiscono i bambini" pensavo scherzasse! 
\includegraphics[height=1em]{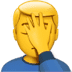}
\includegraphics[height=1em]{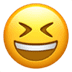}
 & @user When I was little my mum used to say "stay away from gypsies because they kidnap children" I thought she was joking! 
\includegraphics[height=1em]{fig/emoji-facepalm}
\includegraphics[height=1em]{fig/emoji-grinning-squinting-face}
& 1 & 0 & 0 & 0 & 0 \\ \midrule
Fatemi capire, Hitler si era affidato agli astri e ha sbagliato, sterminando 6 milioni di ebrei. E se ci avesse azzeccato invece? & Let me get this straight, Hitler relied on the stars and got it wrong, exterminating 6 million Jews. But what if he got it right? & 1 & 0 & 0 & 1 & 1 \\ \midrule
Ieri 5 morti a \#Parigi per mano di un immigrato convertito all'Islam, sempre ieri a \#Fermo un immigrato girava con un machete pronto a uccidere. Oggi un immigrato ha ucciso due poliziotti a \#Trieste. Serve prudenza e selezione degli ingressi, niente ingenuità. & Yesterday 5 dead in \#Paris at the hands of an immigrant convert to Islam, also yesterday in \#Fermo an immigrant was walking around with a machete ready to kill. Today an immigrant killed two policemen in \#Trieste. We need prudence and selection of entries, no naivety. & 0 & 1 & 0 & 0 & 1 \\ \bottomrule
\end{tabularx}
\caption{Examples of tweets classified by the four ML algorithms\label{tab:example-algo}}
\end{table}

\begin{table}[htbp]
\centering
\begin{tabularx}{\linewidth}{XXcc}
\toprule
Original text of the tweet & English translation & i7 & i6 \\ \midrule
@user @user Facciamo qualcosa per fermare questi coglioni che svendono il paese a negri, islamici e zingari. La feccia dell'umanità. & @user @user Let's do something to stop these assholes selling out the country to niggers, islamics and gypsies. The scum of humanity. & 1 & 1 \\ \midrule
Odio i rom, se potessi gli sparerei.. gentaccia di merda ladra, zozza, Hitler li doveva ammazzare tutti! & I hate the Roma, I'd shoot them if I could... thieving, filthy scum, Hitler should have killed them all! & 1 & 1 \\ \midrule
Confidiamo in Putin che buone mettere in galera Soros e buttare la chiave & We trust Putin who wants to put Soros in jail and throw away the key & 1 & 0 \\ \midrule
Nulla di nuovo! Sono anni che in inghiltera l‘Islam si diffonde come una malattia! Ci sono quartieri in cui la legge e la sharia.. dove dei barbuti vanno a imporre la morale islamica! L‘inghiltera ormai e andata... & Nothing new! Islam has been spreading like a disease in England for years! There are neighbourhoods where the law is sharia... where bearded men go to impose Islamic morality! England is gone... & 0 & 1 \\
\bottomrule
\end{tabularx}
\caption{Examples of tweets annotated as containing the hate indicators: possible violent response (i6) or inciting hatred and violence (i7)\label{tab:possible_violent}}
\end{table}

\end{document}